\documentclass[twocolumn,english,superscriptaddress,citeautoscript,showpacs,preprintnumbers,amsmath,amssymb,prd,floatfix,footinbib]{revtex4-1}

\usepackage{amsmath, amsthm, amssymb, graphicx,bm,amstext,subfigure}
\usepackage{hyperref}

\begin{document}
\title{Mirror moving in quantum vacuum of a massive scalar field}
\author{Qingdi Wang}
\author{William G. Unruh}
\affiliation{Department of Physics and Astronomy, 
The University of British Columbia,
Vancouver, Canada V6T 1Z1}

\begin{abstract}
We present a mirror model moving in the quantum vacuum of a massive scalar field and study its motion under infinitely fluctuating quantum vacuum stress. The model is similar to the one in \cite{PhysRevD.89.085009}, but this time there is no divergent effective mass to weaken the effect of divergent vacuum energy density. We show that this kind of weakening is not necessary. The vacuum friction and strong anticorrelation property of the quantum vacuum are enough to confine the mirror's position fluctuations. This is another example illustrating that while the actual value of the vacuum energy can be physically significant even for nongravitational system, and that its infinite value makes sense, but that its physical effect can be small despite this infinity.
\end{abstract}
\maketitle
\section{introduction}
In \cite{PhysRevD.89.085009} we presented a mirror model with an internal degree of freedom that interacts with a massless scalar field which shows that the value of vacuum energy does have physical significance. Due to vacuum fluctuations, the mirror experiences infinitely fluctuating stress in a magnitude proportional to the value of vacuum energy density. We found that although the infinitely fluctuating quantum vacuum stress provides infinite instantaneous acceleration of the mirror, the mirror's position would be confined in a small region without Brownian-like diffusion. This happens because two special properties of the quantum vacuum: (1) the vacuum friction and (2) the strong anticorrelation of vacuum fluctuations.

The coupling we used in \cite{PhysRevD.89.085009} between derivative of the field and the internal degree of freedom, which is modeled as an harmonic oscillator, has advantages when the field is massless but has the disadvantage that the energy of the oscillator has an ultraviolet divergence. In the calculation processes this divergence provides a divergent effective mass which weakens the effect of the infinities due to the vacuum fluctuation. This gives an impression that our results are highly dependent on this divergent effective mass \cite{PhysRevD.90.087702}. However, in this paper, we present another mirror model without this divergent effective mass weakening and show that this weakening is not necessary. The vacuum friction and strong anticorrelation property of quantum vacuum are enough to confine the mirror's position fluctuations. The model is a mirror with internal degree of freedom interacting with a massive scalar field and with direct coupling to the value of the field instead of its derivative as in \cite{PhysRevD.89.085009}. The details of this model will be given in section \ref{model}. 

Units are chosen throughout such that $c=\hbar=1$.

\section{the mirror model}\label{model}
We consider a mirror model which has an internal harmonic oscillator $q$ with natural frequency $\Omega$.
Its free Lagrangian is $\frac{1}{2} (\dot q ^2+ \Omega^2 q^2)$. Coupled to this
is a massive scalar field $\phi$ with free Lagrangian $\frac{1}{2}\int(\dot\phi^2 -
{\phi'}^2-m^2\phi^2)dx$, where the dot $\dot{•}$ and the prime $'$ denote the time and space derivatives respectively. We will consider the simplest coupling, the one between the internal degree of freedom $q$ and $\phi$ given by $\int \epsilon q \phi \delta(x) dx$.
The action of the whole system is, thus,
\begin{equation}\begin{split}\label{action}
S=&\frac{1}{2}\iint\left(\dot{\phi}^2-\phi'^2-m^2\phi^2\right)dtdx\\
&+\frac{1}{2}\int\left(\dot{q}^2-\Omega^2q^2\right)dt\\
&+\epsilon\int q(t)\phi(t,0)dt.
\end{split}\end{equation}
Varying \eqref{action} with respect to the field $\phi$ and the internal degree of freedom $q$ give the equations of motion:
\begin{equation}\label{eom1}
\ddot{\phi}-\phi''+m^2\phi=\epsilon q\delta(x),
\end{equation}
\begin{equation}\label{q equation}
\ddot{q}+\Omega^2q=\epsilon\phi(t,0).
\end{equation}

For the massless case $m=0$, this model is the same with the model $(2.1)$ in \cite{PhysRevA.87.043832}. Unfortunately, in this massless case the internal degree of freedom, $q$, is unstable, which was not mentioned by the authors of \cite{PhysRevA.87.043832}. To see this, first notice that the solution for $\phi$ is of the form
\begin{equation}\label{phi solution}
\phi=\phi_0+\frac{\epsilon}{2}\int^{t-|x|}dt'q(t'),
\end{equation}
where $\phi_0$ satisfies the homogeneous field equation
\begin{equation}\label{massless homogeneous field equation}
\ddot{\phi_0}-\phi_0''=0.
\end{equation}
Then inserting \eqref{phi solution} into \eqref{q equation} and taking time derivative gives
\begin{equation}\label{q equation1}
\dddot{q}+\Omega^2\dot{q}-\frac{\epsilon^2}{2}q=\epsilon\dot{\phi_0}(t,0).
\end{equation}
The characteristic equation of the above ordinary differential equation \eqref{q equation1}
\begin{equation}\label{characteristic equation}
\lambda^3+\Omega^2\lambda-\epsilon^2/2=0
\end{equation}
has a positive root, which means that the internal degree of freedom $q$ is unstable. However, if the field is massive enough, the system becomes stable. In the following, we will give the solution of the system and explain why it is stable if the field is massive enough.

By doing Fourier transform of the field $\phi$ and the internal degree of freedom $q$, we find that there are two different classes of mode solutions. The first class is a continuum class of modes which go as $e^{-i\omega t}$ with frequencies $\omega=+\sqrt{k^2+m^2}\geq m$. These have an ingoing field which resonantly excites the $q$ mode and is radiated:
\begin{eqnarray}\label{mode solution 1}
\phi_k(t,x) &=&c_1 \left(e^{i k x}-\frac{\epsilon^2a_ke^{i|k||x|}}{2i|k|\left(-\omega^2+\Omega^2\right)+\epsilon^2}\right)e^{-i\omega t},\\
q_k(t) &=& c_1\frac{2i|k|\epsilon}{2i|k|(-\omega^2+\Omega^2)+\epsilon^2} e^{-i\omega t}, \label{mode solution 2}
\end{eqnarray}
where $c_1$ is a constant and $-\infty<k<+\infty$.

The second class is an isolated and localized mode which goes as $e^{-i\hat{\omega}t}$:
\begin{eqnarray}\label{mode solution 3}
\phi_{\kappa}(t,x) &=&c_2e^{-\kappa|x|}e^{-i\hat{\omega}t},\\
q_{\kappa}(t) &=&c_2\frac{2\kappa}{\epsilon}e^{-i\hat{\omega}t},\label{mode solution 4}
\end{eqnarray}
where $c_2$ is a constant and $\kappa$ is one of the three roots of the cubic equation:
\begin{equation}\label{cubic equation}
f(\kappa)=\kappa^3+(\Omega^2-m^2)\kappa-\epsilon^2/2=0,
\end{equation}
and the frequency $\hat{\omega}$ is related to $\kappa$ by $\hat{\omega}=\sqrt{m^2-\kappa^2}$.

To make sure that the mode solution\eqref{mode solution 3} is well defined we require that the real part of $\kappa$ is positive, since otherwise \eqref{mode solution 3} will blow up at spatial infinity. In the following we analyze which roots of \eqref{cubic equation} satisfy this requirement.

First, since $f(0)=-\epsilon^2/2<0$ and $f(\kappa)\to+\infty$, as $\kappa\to+\infty$, the cubic equation \eqref{cubic equation} always has one real positive root, which is denoted by $\kappa_1$. For the other two roots, which are denoted by $\kappa_2$ and $\kappa_3$, there are two different cases.

Case I: $\kappa_2$ and $\kappa_3$ are complex conjugate. In this case, because the coefficient of the quadratic term of $f(\kappa)$ is $0$, we have the relation between the three roots:
\begin{equation}
\kappa_1+\kappa_2+\kappa_3=0.
\end{equation}
Because $\kappa_1$ is real and positive, the real part of $\kappa_2$ and $\kappa_3$ must be negative.

Case II: $\kappa_2$ and $\kappa_3$ are both real. In this case $f(\kappa)$ has two equally spaced stationary points at $\kappa=\pm\sqrt{(m^2-\Omega^2)/3}$. Also at these two points $f(-\sqrt{(m^2-\Omega^2)/3})\geq 0$ and $f(+\sqrt{(m^2-\Omega^2)/3})\leq -\epsilon^2/2$. Then from the property of continuity of $f(\kappa)$ we know that $\kappa_2$ and $\kappa_3$ are both negative.

Thus in summary, $\kappa_1$ is the only root that satisfies the requirement of having a positive real part and thus, the $\kappa$ in \eqref{mode solution 3} and \eqref{mode solution 4} can only be $\kappa_1$.

Next we need to consider the stability of \eqref{mode solution 3} and \eqref{mode solution 4} in time. For the $e^{\pm i\hat{\omega} t}$ be stable, $\hat{\omega}$ must be real. Since $\hat{\omega}$ is related to $\kappa_1$ by $\hat{\omega}=\sqrt{m^2-\kappa_1^2}$, we then must have $m>\kappa_1$. This requirement is equivalent to the condition that $f(m)>f(\kappa_1)=0$, which leads to $m>\frac{\epsilon^2}{2\Omega^2}$. 

In the following we only consider the stable case with large enough $m$.

\section{quantization of the system}
The system can be canonically quantized by standard procedure \cite{quantization}. First, we define the inner product of the solutions of \eqref{eom1} and \eqref{q equation} by
\begin{equation}\begin{split}\label{inner product}
\left((\phi_1, q_1),(\phi_2, q_2)\right)=&-i\int_{-\infty}^{\infty}dx\left(\phi_1\partial_t\phi_2^*-\phi_2^*\partial_t\phi_1\right)\\
&-i\left(q_1\partial_tq_2^*-q_2^*\partial_tq_1\right).
\end{split}\end{equation}

The mode solutions \eqref{mode solution 1}, \eqref{mode solution 2}, \eqref{mode solution 3} and \eqref{mode solution 4} are orthogonal to each other under the above inner product definition. We can further normalize these mode solutions by the following conditions
\begin{eqnarray}
\left((\phi_k, q_k),(\phi_{k'}, q_{k'})\right)&=&\delta(k-k'),\\
\left((\phi_{\kappa}, q_{\kappa}),(\phi_{\kappa}, q_{\kappa})\right)&=&1.
\end{eqnarray}

Then the system can be quantized by expanding the field operator $\phi$ and the internal degree of freedom $q$ as the sum of normalized mode solutions \eqref{mode solution 1}, \eqref{mode solution 2}, \eqref{mode solution 3} and \eqref{mode solution 4} with creation and annihilation operators as coefficients:
\begin{eqnarray}
\phi&=&\int_{-\infty}^{\infty}dk(a_k\phi_k+a_k^{\dag}\phi_k^*)+(A\phi_{\kappa}+A^{\dag}\phi_{\kappa}^*),\\
q&=&\int_{-\infty}^{\infty}dk(a_kq_k+a_k^{\dag}q_k^*)+(Aq_{\kappa}+A^{\dag}q_{\kappa}^*).
\end{eqnarray}
The creation and annihilation operators would satisfy the standard commutation relations:
\begin{eqnarray}
\left[a_k, a_{k'}^{\dag}\right]&=&\delta(k-k')\\
\left[A, A^{\dag}\right]&=&1.
\end{eqnarray}

In the following we summarize the result of the above quantization method.

The field operator $\phi$ can be decomposed into three parts:
\begin{equation}\label{phi decomposition}
\phi(t,x)=\phi_0(t,x)+\phi_1(t,|x|)+\phi_2(t,|x|).
\end{equation}

The $\phi_0$ in \eqref{phi decomposition}, which satisfies
\begin{equation}
\ddot{\phi_0}-\phi_0''+m^2\phi_0=0,
\end{equation}
includes all the free field modes $e^{-i(\omega t-k x)}$. It is expanded in terms of annihilation and creation operators:
\begin{equation}\label{phi0 expression}
\phi_0(t,x)=\int_{-\infty}^{\infty}dk\frac{1}{\sqrt{4\pi\omega}}\left[a_ke^{-i\left(\omega t-kx\right)}+a_k^{\dag}e^{i\left(\omega t-kx\right)}\right].
\end{equation}

The $\phi_1$ in \eqref{phi decomposition} includes all the modes $e^{-i(\omega t-|k||x|)}$:
\begin{equation}\label{phi1 expression}\begin{split}
\phi_1(t,|x|)=&-\int_{-\infty}^{\infty}dk\frac{1}{\sqrt{4\pi\omega}}\Bigg[\frac{\epsilon^2a_ke^{-i\left(\omega t-|k||x|\right)}}{2i|k|\left(-\omega^2+\Omega^2\right)+\epsilon^2}\\
&+\frac{\epsilon^2a_k^{\dag}e^{i\left(\omega t-|k||x|\right)}}{-2i|k|\left(-\omega^2+\Omega^2\right)+\epsilon^2}\Bigg].
\end{split}\end{equation}

The $\phi_2$ in \eqref{phi decomposition} includes the single mode $e^{-i\hat{\omega}t-\kappa |x|}$:
\begin{equation}\label{phi2 expression}
\phi_2(t,|x|)=\sqrt{\frac{\kappa\epsilon^2}{2\hat{\omega}\left(4\kappa^3+\epsilon^2\right)}}\left(Ae^{-i\hat{\omega t}}+A^{\dag}e^{i\hat{\omega t}}\right)e^{-\kappa|x|}.
\end{equation}

The internal oscillator $q$ is expanded as:
\begin{equation}\begin{split}\label{q expansion}
q(t)=&\int_{-\infty}^{\infty}dk\frac{1}{\sqrt{4\pi\omega}}\bigg(\frac{2i|k|\epsilon a_ke^{-i\omega t}}{2i|k|(-\omega^2+\Omega^2)+\epsilon^2}\\
&+\frac{-2i|k|\epsilon a_k^{\dag}e^{i\omega t}}{-2i|k|(-\omega^2+\Omega^2)+\epsilon^2}\bigg)\\
&+\sqrt{\frac{2\kappa^3}{\hat{\omega}(4\kappa^3+\epsilon^2)}}\left(Ae^{-i\hat{\omega t}}+A^{\dag}e^{i\hat{\omega t}}\right).
\end{split}\end{equation}
We can see from the above expression \eqref{q expansion} that, for large $k$, the integrand
\begin{equation}
q(k)\propto \frac{1}{k^{5/2}}(a_k+a_k^{\dag}),\quad \dot{q}(k)\propto \frac{1}{k^{3/2}}(a_k-a_k^{\dag}).
\end{equation}
Then the expectation value of internal energy of the mirror, for large $k$, goes as
\begin{equation}
\left\langle\frac{1}{2}(\dot{q}^2+\Omega^2q^2)\right\rangle\sim\int dk\frac{1}{k^3}<\infty.
\end{equation}
Therefore, unlike the mirror in \cite{PhysRevD.89.085009}, the integrand here drops fast enough in response to high frequency modes that the motion of $q$ only adds a finite amount of energy to the mirror's effective mass.
\section{The force exerted on the mirror}
As in \cite{PhysRevD.89.085009}, the force exerted on the mirror is defined as the pressure difference from both sides:
\begin{equation}\label{force definition}
F(t)=\lim_{x\to 0^+}\left(T^{11}(x_-)-T^{11}(x_+)\right),
\end{equation}
where $x_+=(t,x)$ and $x_-=(t,-x) (x\geq 0)$ are two spacetime points which are symmetrically located on the two sides of the mirror and $T^{11}$ is the space-space component of stress-energy tensor of type $(2,0)$ of the field $\phi$:
\begin{equation}\label{stress energy}
T^{11}(t,x)=\frac{1}{2}\left(\dot{\phi}^2(t,x)+\phi'^2(t,x)-m^2\phi^2\right).
\end{equation}
Inserting \eqref{phi decomposition} and \eqref{stress energy} into \eqref{force definition} and noticing that when $x\to 0^+$, due to continuity of the field $\phi$, only the following terms survive:
\begin{equation}\label{force}
F(t)=2\left\{\phi_0'(t,0)\left(\phi_1'\left(t,0\right)+\phi_2'\left(t,0\right)\right)\right\},
\end{equation}
where the curly bracket $\{\}$ is the symmetric product which is defined as
\begin{equation}
\{AB\}=\frac{1}{2}(AB+BA).
\end{equation}

The expectation value of the force \eqref{force} is zero when evaluated in the incoming vacuum state. However, the fluctuation of this force in the vacuum state
\begin{equation}\label{force fluctuation definition}
\sigma_F(t)=\left\langle F^2(t)\right\rangle-\left\langle F(t)\right\rangle^2
\end{equation}
is not. In fact, if we insert \eqref{force} into \eqref{force fluctuation definition} and then take Wick's expansion, we get
\begin{equation}
\sigma_F(t)=4\left\langle\phi_0'^2(t,0)\right\rangle\left\langle\left(\phi_1'\left(t,0\right)+\phi_2'\left(t,0\right)\right)^2\right\rangle.
\end{equation}
Noticing that, in the $1+1$ dimension, the term $\left\langle\phi_0'^2\right\rangle=\frac{1}{2}\left\langle\dot{\phi_0}^2+\phi_0'^2-m^2\phi^2\right\rangle=\left\langle T^{11}\right\rangle$, we obtain that the fluctuation of force exerted on the mirror is proportional to the expectation value of Minkowski vacuum stress:
\begin{equation}\label{force fluctuation}
\sigma_F(t)\sim4\left\langle\left(\phi_1'\left(t,0\right)+\phi_2'\left(t,0\right)\right)^2\right\rangle\left\langle T^{11}\right\rangle.
\end{equation}

The difference between vacuum stress $T^{11}$ and the vacuum energy density $T^{00}$, which is defined as
\begin{equation}
T^{00}(t,x)=\frac{1}{2}\left(\dot{\phi}^2(t,x)+\phi'^2(t,x)+m^2\phi^2\right),
\end{equation}
is $m^2\phi^2$. It is only the logarithmic divergence compared to the quadratic divergence for the expectation value of vacuum energy density in $1+1$ dimension; thus, the divergence of the fluctuation of force exerted on the mirror is in the same order as the divergence of the vacuum energy density.

\section{vacuum friction due to Doppler effect}
\begin{figure}
\centering
\includegraphics[scale=0.5]{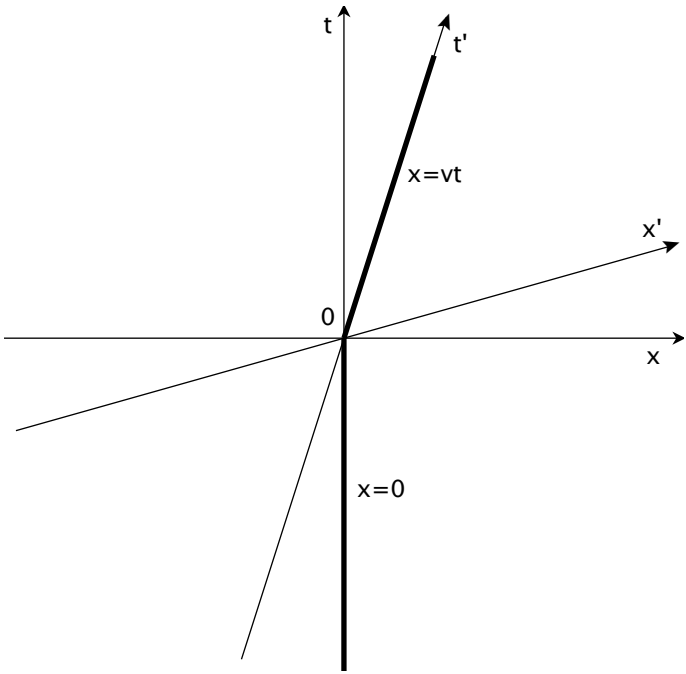}
\caption{\label{jump}The trajectory for a mirror that initially stays at rest and then jumps to move with a constant velocity $v$ at time $t=0$.}
\end{figure}

Similar to \cite{PhysRevD.89.085009}, if the mirror initially stays at rest, it would experience a friction force when it starts to move. The friction force arises from the Doppler shift of the reflected vacuum modes due to the changing velocity of the mirror. In this section we consider that the mirror initially has been at rest for a long time. It then starts to move with a constant velocity $v$ (see FIG. \ref{jump})). We calculate the expectation value of the vacuum friction at the jump point $t=0$. We will consider everything in the mirror's instantaneous rest frame.

For the trajectory shown in FIG. \ref{jump}, when $t<0$, the mirror's rest frame is $(t,x)$ coordinate system and the field $\phi_0$ is expanded as the sum of positive frequency modes $\frac{e^{-i(\omega t-kx)}}{\sqrt{4\pi\omega}}$ with coefficients $a_k$ and negative frequency modes $\frac{e^{+i(\omega t-kx)}}{\sqrt{4\pi\omega}}$ with coefficients $a_k^{\dag}$ (see \eqref{phi0 expression}). When $t\geq 0$, the mirror's rest frame is $(t',x')$ coordinate system (see FIG (\ref{jump})) and the same field $\phi_0$ is expanded as:
\begin{equation}\label{phi0 transform}
\phi_0(t',x')=\int_{-\infty}^{\infty}\frac{dk'}{\sqrt{4\pi\omega'}}\left[b_{k'}e^{-i\left(\omega' t'-k'x'\right)}+b_{k'}^{\dag}e^{i\left(\omega' t'-k'x'\right)}\right],
\end{equation}
where the $\omega'$, $k'$ in $(t',x')$ coordinate system are Doppler shifted from the $\omega$, $k$ in $(t,x)$ system to:
\begin{eqnarray}\label{transform relation1}
\omega'&=&\gamma(\omega-kv),\\
\label{transform relation2}
k'&=&\gamma(k-\omega v),
\end{eqnarray}
and correspondingly, the operator coefficients $b_{k'}$ and $a_k$ are related by
\begin{equation}\label{operator transform}
b_{k'}=\left(\gamma\left(1+k'v/\omega'\right)\right)^{1/2}a_k,
\end{equation}
where $\gamma=\frac{1}{\sqrt{1-v^2}}$ is the Lorentz factor.

Due to Lorentz invariance, the expression of the friction force in $(t',x')$ frame is exactly the same form as in $(t,x)$ frame \eqref{force}. The only difference is that the spatial derivative $'$ is now with respect to $|x'|$ instead of $|x|$. Therefore, the friction force at the jumping point $t=0$ is
\begin{equation}\label{jump force}
F(0)=2\left\{\phi_0'(0,0)\left(\phi_1'\left(0,0\right)+\phi_2'\left(0,0\right)\right)\right\},
\end{equation}
where
\begin{equation}\label{phi1 in tx frame}
\phi_1'(0,0)=-\epsilon^2\int_{-\infty}^{\infty}\frac{dk}{\sqrt{4\pi\omega}}\left[\frac{i|k|a_k}{2i|k|(-\omega^2+\Omega^2)+\epsilon^2}+c.c\right].
\end{equation}
Transforming the above expression \eqref{phi1 in tx frame} for $\phi_1'(0,0)$ from the $(t,x)$ frame to the $(t',x')$ frame by using \eqref{transform relation1}, \eqref{transform relation2} and \eqref{operator transform} gives
\begin{widetext}
\begin{equation}\label{phi1  t'x'}
\phi_1'(0,0)=-\epsilon^2\int_{-\infty}^{\infty}\frac{dk'}{\sqrt{4\pi\omega'}}\left[\frac{i\gamma|k'+\omega'v|b_{k'}}{2i\gamma |k'+\omega'v|(-\gamma^2(\omega'+k'v)^2+\Omega^2)+\epsilon^2}+c.c\right].
\end{equation}
\end{widetext}
Since the operators $A$, $A^{\dag}$ and $a_k$, $a_k^{\dag}$ commute, the expectation value of the second term, $\phi_0'\phi_2'$, in \eqref{jump force} is zero. Thus when taking the expectation value of the force at $t=0$, only the first term $\phi_0'\phi_1'$ survives. Inserting \eqref{phi0 transform} and \eqref{phi1  t'x'} into \eqref{jump force} and taking the expectation value in the vacuum state gives the friction force:
\begin{equation}\label{force result}
F=\left(\frac{\epsilon^4}{\pi}\int_0^{+\infty}dk\frac{k}{4k^2(\Omega^2-\omega^2)^2+\epsilon^4}\right)\gamma v.
\end{equation}

\section{fluctuating motion of the mirror}\label{fluctuating motion of the mirror}
In this section, we investigate how the mirror moves under the infinitely fluctuating quantum vacuum stress \eqref{force fluctuation}. Analogues to the motion of the mirror in \cite{PhysRevD.89.085009}, the nonrelativistic equation of motion of the mirror here can be modeled as
\begin{equation}\label{non relavitistic emo}
dv/dt+\beta(v)v=F/M,
\end{equation}
where $M$ is mass of the mirror and $\beta$ is the damping coefficient. From \eqref{force result} we know that $\beta$ is velocity dependent with the property that
\begin{equation}\label{asymptotic condition}
\lim_{v\to 1}\beta=+\infty.
\end{equation}

For simplicity, we first assume that $\beta$ is constant and then the solution of the mirror's velocity and position with initial conditions $X(0)=0, v(0)=0$ can be expressed as
\begin{equation}
v(t)=\frac{1}{M}e^{-\beta t}\int_0^tdt'e^{\beta t'}F(t'),
\end{equation}
\begin{equation}
X(t)=\frac{1}{M}\int_0^tdt'e^{-\beta t'}\int_0^{t'}dt''e^{\beta t''}F(t'').
\end{equation}
Then we can directly calculate the fluctuation of the mirror's velocity
\begin{eqnarray}\label{velocity fluctuation}
&&\sigma_v(t)=\left\langle v(t)^2\right\rangle-\left\langle v(t)\right\rangle^2\\
~~~~&&=\frac{1}{M^2}e^{-2\beta t}\int_0^t\int_0^tdt_1dt_2e^{\beta(t_1+t_2)}Corr(F(t_1),F(t_2)),
\nonumber\end{eqnarray}
and the fluctuation of the mirror's position
\begin{widetext}
\begin{equation}\label{position fluctuation}
\sigma_X(t)=\left\langle X(t)^2\right\rangle-\left\langle X(t)\right\rangle^2=\frac{1}{M^2}\int_0^tdt_1e^{-\beta t_1}\int_0^{t_1}dt_2e^{\beta t_2}\cdot\int_0^tdt_3e^{-\beta t_3}\int_0^{t_3}dt_4e^{\beta t_4}Corr(F(t_2),F(t_4)),
\end{equation}
where the correlation function
\begin{equation}
 Corr(F(t_1),F(t_2))=\left\langle F(t_1)F(t_2)\right\rangle-\left\langle F(t_1)\right\rangle\left\langle F(t_2)\right\rangle
\end{equation}
can be obtained by first inserting \eqref{phi0 expression}, \eqref{phi1 expression} and \eqref{phi2 expression} into \eqref{force} and then taking a Wick's expansion. The result is
\begin{equation}\label{correlation function}
Corr(F(t_1), F(t_2))=\frac{\epsilon^4}{16\pi^2}\int_{-\infty}^{\infty}dk\frac{k^2}{\omega}e^{-i\omega(t_1-t_2)}\cdot\Bigg(\int_{-\infty}^{\infty}dk'\frac{k'^2e^{-i\omega'(t_1-t_2)}}{\omega'\left(k'^2\left(-\omega'^2+\Omega^2\right)^2+\epsilon^4/4\right)}+\frac{8\pi\kappa^3e^{-i\hat{\omega}(t_1-t_2)}}{\hat{\omega}\left(4\kappa^3+\epsilon^2\right)\epsilon^2}\Bigg).
\end{equation}
Inserting \eqref{correlation function} into \eqref{velocity fluctuation} we obtain
\begin{equation}\label{mean squared velocity}\begin{split}
\sigma_v(t)=\frac{\epsilon^4}{16\pi^2M^2}\int_{-\Lambda}^{\Lambda}dk\frac{k^2}{\omega}&\cdot\Bigg(\int_{-\infty}^{\infty}dk'\frac{k'^2\left(1-2e^{-\beta t}\cos(\omega+\omega')t+e^{-2\beta t}\right)}{\omega'\left(k'^2\left(-\omega'^2+\Omega^2\right)^2+\epsilon^4/4\right)\left(\beta^2+\left(\omega+\omega'\right)^2\right)}\\
&+\frac{8\pi\kappa^3\left(1-2e^{-\beta t}\cos(\omega+\hat{\omega})t+e^{-2\beta t}\right)}{\hat{\omega}\left(4\kappa^3+\epsilon^2\right)\epsilon^2\left(\beta^2+\left(\omega+\hat{\omega}\right)^2\right)}\Bigg),
\end{split}\end{equation}
where $\Lambda$ is high frequency cutoff. Similarly, inserting \eqref{correlation function} into \eqref{position fluctuation} we obtain
\begin{equation}\label{mean squared position}
\begin{split}
\sigma_X(t)=\frac{\epsilon^4}{16\pi^2M^2}\int_{-\Lambda}^{\Lambda}dk\frac{k^2}{\omega}&\cdot\Bigg(\int_{-\infty}^{\infty}dk'\frac{k'^2\left[\frac{1}{\beta^2}(1-e^{-\beta t})^2+\frac{4\sin^2(\frac{\omega+\omega'}{2}t)}{(\omega+\omega')^2}-\frac{1}{\beta}(1-e^{-\beta t})\frac{2\sin(\omega+\omega')t}{\omega+\omega'}\right]}{\omega'\left(k'^2\left(-\omega'^2+\Omega^2\right)^2+\epsilon^4/4\right)\left(\beta^2+\left(\omega+\omega'\right)^2\right)}\\
&+\frac{8\pi\kappa^3\left[\frac{1}{\beta^2}(1-e^{-\beta t})^2+\frac{4\sin^2(\frac{\omega+\hat{\omega}}{2}t)}{(\omega+\hat{\omega})^2}-\frac{1}{\beta}(1-e^{-\beta t})\frac{2\sin(\omega+\hat{\omega})t}{\omega+\hat{\omega}}\right]}{\hat{\omega}\left(4\kappa^3+\epsilon^2\right)\epsilon^2\left(\beta^2+\left(\omega+\hat{\omega}\right)^2\right)}\Bigg)
\end{split}
\end{equation}
\end{widetext}
As $\Lambda\to\infty$ and $t\to\infty$, $\sigma_X$ and $\sigma_v$ have a simple relation:
\begin{equation}\label{relation}
\sigma_X\sim\frac{1}{\beta^2}\sigma_v.
\end{equation}

The mean squared velocity $\sigma_v$ is logarithmically divergent, which is a very slow divergence, as the high frequency cutoff $\Lambda$ goes to infinity.  Unlike in \cite{PhysRevD.89.085009}, there is no logarithmically divergent effective mass to cancel the divergence in the mean squared velocity\eqref{mean squared velocity} as the cutoff $\Lambda$ go to infinity. However, notice that even in \cite{PhysRevD.89.085009}, we have $\sigma_v=2$, which is still faster than the speed of light. For the same reason as in \cite{PhysRevD.89.085009}, this unphysical result comes from the small constant damping coefficient $\beta$ assumption we made in the beginning. Actually from the expression for the friction force \eqref{force result} we see that the damping coefficient is monotonically increasing as the velocity increases. When the mirror's velocity approaches 1, the damping coefficient $\beta$ (see \eqref{force result} and \eqref{asymptotic condition}) goes to infinity to make sure that the mirror's velocity never reaches the speed of light. If we further fully consider the relativistic effect, the increased mirror's ``relativistic mass'' would just make the result even smaller. Therefore, we can conclude that the mean squared velocity
\begin{equation}
\sigma_v<1.
\end{equation}

Thus from the relation \eqref{relation} we obtain that the mean squared position
\begin{equation}
\sigma_X<\frac{1}{\beta^2}.
\end{equation}
Since the mirror's speed approaches the light speed $1$, we then get from the property \eqref{asymptotic condition} that the damping coefficient $\beta$ would be, on average, very large, which means that while the force on the mirror and its velocity undergo wild fluctuations, its position fluctuations would again be expected to be confined in a small region.

Notice that we derived the relation \eqref{relation} based on the nonrelativistic equation of motion \eqref{non relavitistic emo} and the assumption that the damping coefficient is constant. This relation might not valid if we fully consider the relativistic effect and the nonconstancy of the damping, but the conclusion should not be affected. That is because our conclusion does not essentially rely on \eqref{relation}. The key point is that, in the long time limit, the magnitude of the fluctuations of the velocity and position are time independent, i.e. it does not grow with time. This nondiffusing property is guaranteed by the strong anticorrelation of the the quantum fluctuation, which is manifest in \eqref{mean squared velocity} and \eqref{mean squared position}. When considering the exact relativistic effect and the monotonically increasing damping as velocity increases, the anticorrelation of the quantum fluctuation is still there, and the only difference is that the mirror would be more difficult to move, which would make the magnitude of the mirror's fluctuating motion even smaller. Thus the mirror would still be confined but moving back and forth with a speed close to the light speed due to the infinitely fluctuating quantum vacuum. The fluctuation time scale would be on the order of cutoff time scale $1/\Lambda$ and the range of the mirror's fluctuating motion would be confined in
\begin{equation}
\Delta X\sim \frac{1}{\Lambda}.
\end{equation}
This goes to $0$ as we take the high frequency cutoff $\Lambda$ to infinity; i.e., we would expect the fluctuating forces not to move the mirror.

\section{conclusions}
We present a mirror moving in quantum vacuum of a massive scalar field which is similar to the massless one in \cite{PhysRevD.89.085009}. The finite mass allows a stable nonderivative coupling to the coordinate of the oscillator without a divergent self-energy. In both cases the field exerts a fluctuating force on the mirror in a magnitude proportional to the infinite value of the vacuum energy density. The main difference from \cite{PhysRevD.89.085009} is that we are using a different coupling and, by necessity, a massive field. In the calculation process of the mirror model of \cite{PhysRevD.89.085009}, there exists a divergent effective mass to weaken the effect of the infinite vacuum fluctuations. However, this weakening does not exist in the model we used in this paper. The vacuum friction and strong anticorrelation property of quantum vacuum are enough to confine the mirror's position fluctuations.

This is another example illustrating that while the actual value of vacuum energy can be physically significant even for a nongravitational system, and that its infinite value makes sense, but that its physical effect can be small despite this infinity.

\section*{Acknowledgments}
W.G.U would like to thank the Canadian Institute for Advanced Research (CIFAR), the Natural Sciences and Engineering Research Council of Canada (NSERC), and the John Templeton Foundation for their support of this research. Qingdi Wang would like to thank UBC for their support of studies during this work through Faculty of Science Graduate Award.

\bibliographystyle{unsrt}
\bibliography{vacuum}

\end{document}